\documentclass[prl,aps,twocolumn]{revtex4}

\usepackage{graphics}

\begin{document}

\title{One-time pad booster for Internet}

\author{Geraldo A. Barbosa$^*$}
\affiliation{QuantaSec -- Research in Quantum Cryptography Ltd., 1558 Portugal Ave., Belo Horizonte MG 31550-000 Brazil. }

\date{11 April 2007}
\begin{abstract}

One-time pad encrypted files can be sent through Internet channels using current Internet protocols. However, the need for
renewing shared secret keys make this method unpractical. This work shows how users can use a fast physical random generator
based on fluctuations of a light field and the Internet channel to directly boost key renewals. The transmitted signals are
deterministic but carries imprinted noise that cannot be eliminated by the attacker. Thus, a one-time pad for Internet can
be made practical. Security is achieved without third parties and not relying on the difficulty of factoring numbers in
primes. An informational fragility to be avoided is discussed. Information-theoretic analysis is presented and bounds for
secure operation are determined.


\end{abstract}

\maketitle

\newcommand{\be}{\begin{equation}}
\newcommand{\ee}{\end{equation}}
\newcommand{\bea}{\begin{eqnarray}}
\newcommand{\eea}{\end{eqnarray}}


Unconditionally secure one-time pad encryption \cite{vernam} has not find wide applicability in modern communications. The
difficult for users to share long streams of secret keys beforehand has been an unsurmountable barrier preventing widespread
use of one-time pad systems. Even beginning with a start sequence of shared secret keys, no amplification method to obtain
new key sequences or key ``refreshing'' is available. This work proposes a practical solution for this problem and discusses
its own limitations.

Assume that (statistical) physical noise ${\bf n}=n_1,n_2,...$ has
been added to a message bit sequence ${\bf X}=x_1,x_2,...$ according
to some rule $f_j(x_j,n_j)$ giving ${\bf
Y}=f_1(x_1,n_1),f_2(x_2,n_2),...\:$ (Whenever binary physical
signals are implied, use $f_j(x_j,n_j)$ will represent $f_j =
\oplus$ ($=$addition mod2)). When analog physical signals are made
discrete by analog-to-digital converters,  a sum of a binary signal
onto a discrete set will be assumed). The addition process is
performed at the emitter station and ${\bf Y}$ becomes a binary file
carrying the recorded noise. ${\bf Y}$ is sent from user A to user B
(or from B to A) through an insecure channel. The amount of noise is
assumed high and such that without any knowledge beyond ${\bf Y}$,
neither B (or A) or an attacker E could extract the sequence ${\bf
X}$ with a probability $P$ better than the guessing level of
$P=(1/2)^N$, where $N$ is the number of bits.

Assuming that A and B share some knowledge beforehand, the amount of information between A (or B) and E differs. Can this
information asymmetry be used by A and B to share secure information over the Internet?  It will be shown that if A and B
start sharing a secret key sequence ${\bf K}_0$ they may end up with a practical new key sequence ${\bf K} \gg {\bf K}_0$.
The security of this new sequence is discussed including an avoidable fragility for a-posteriori attack with a
known-plaintext attack.  Within bounds to be demonstrated, this makes one-time pad encryption practical for fast Internet
communications (data, image or sound). It should be emphasized that being practical does not imply that ${\bf K}_0$ or the
new keys have to be open to the attacker after transmission. These keys have to be kept secret as long as encrypted messages
have to be protected, as in a strict one-time pad. The system gives users A and B direct control to guarantee secure
communication without use of third parties or certificates. Some may think of the method as an extra protective layer to the
current Internet encryption protocols. The system operates on top of all IP layers and does not disturb current protocols in
use by Internet providers. Anyway, one should emphasize that the proposed method relies on security created by physical
noise and not just on mathematical complexities such as the difficulty of factoring numbers in primes. This way, its
security level does not depend on advances in algorithms or computation.

Random events of physical origin cannot be deterministically
predicted and sometimes are classified in classical or quantum
 events. Some take the point of view that a recorded classical random event is
just the record of a single realization among all the possible
quantum trajectories possible \cite{belavkin}. These classifications
belong to a philosophical nature, and are not relevant to the
practical aspects to be discussed here. However, what  should be
emphasized is that physical noise is completely different from
pseudo noise generated in a deterministic process (e.g. hardware
stream ciphers) because despite any complexity introduced, the
deterministic generation mechanism can be searched, eventually
discovered and used by the attacker.



Before introducing the communication protocol to be used, one should
discuss the superposition of physical signals to deterministic
binary signals. Any signal transmitted over Internet is physically
prepared to be compatible with the channel being used. This way,
e.g., voltage levels $V_0$ and $V_1$ in a computer may represent
bits. These values may be understood as the simple encoding
\begin{eqnarray}
V^{(0)}\Rightarrow\left\{
\begin{array}{c}
V_0 \rightarrow \mbox{bit} \:0\\V_1 \rightarrow \mbox{bit}\:1 \end{array} \right.
\end{eqnarray}
Technical noise, e.g. electrical noise, in bit levels $V_0$ and $V_1$ are assumed low. Also, channel noise are assumed with
a modest level. Errors caused by these noises are assumed to be possibly corrected by classical error-correction codes.
Anyway, the end user is supposed to receive the bit sequence ${\bf X}$
 (prepared by a sequence of $V_0$ and $V_1$) as determined by the
sender. If one of these deterministic binary signals $x_j$ is repeated over the channel, e.g. $x_1=x$ and $x_2=x$, one has
the known property $x_1\oplus x_2=0$. This property has to be compared to cases where a non-negligible amount of physical
noise $n_j$ (in analog or a discrete  form) has been added to each emission. Writing $y_1=f_1(x_1,n_1)=f_1(x,n_1)$ and
$y_2=f_2(x_2,n_2)=f_2(x,n_2)$ one has $f(y_1,y_2)=$
 neither $0$ or $1$ in general.
This difference from the former case where $x_1\oplus x_2=0$
emphasizes the uncontrollable effect of the noise.

The $V^{(0)}$ encoding shown above allows binary values $V_0$ and
$V_1$ to represent bits 0 and 1, respectively. These values are
assumed to be determined without ambiguity. Instead of this unique
encoding  consider that {\em two} distinct encodings can be used to
represent bits 0 and 1: Either $V^{(0)}$ over which $x^{(0)}_0$ and
$x^{(0)}_1$ represent the two bits 0 and 1 respectively, or
$V^{(1)}$, over which $x^{(1)}_1=x^{(0)}_0+\epsilon$ and
$x^{(1)}_0=x^{(0)}_1+\epsilon$ ($\epsilon \ll 1$) represent the two
bits 1 or 0 (in a different order from the former assignment). These
encodings represent physical signals as, for example, phase signals.

Assume noiseless transmission signals but where noise $n_j$ has been
introduced or added to each $j^{\mbox{\tiny th}}$ bit sent (This is
equivalent to noiseless signals in a noisy channel). Consider that
the user does not know which encoding $V^{(0)}$ or $V^{(1)}$ was
used. With a noise level $n_{j}$ superposed to signals in $V^{(0)}$
or $V^{(1)}$ and if $|x^{0}_0-x^{1}_0|\gg n_j \gg \epsilon$, one
cannot distinguish between signals 0 and 1 in $V^{(0)}$ and
$V^{(1)}=V^{(0)}+\epsilon$ but one knows easily that a signal
belongs either to the set $(0 \:\:\mbox{in}\: V^{(0)}\: \mbox{ or}
\:1 \mbox{ in}\:  \:V^{(1)})$ {\bf or} to the set $(1 \:\mbox{ in}\:
V^{(0)}\: \mbox{or}\: 0 \:\mbox{in}\:  V^{(1)})$. Also note that
once the encoding used is known, there is no question to identify
between $x_j$ and $x_j+\epsilon$. In this case, it is
straightforward to determine a bit 0 or 1 because values in a single
encoding are widely separated and, therefore, distinguishable. One
may say that without information on the encoding used, the bit
values cannot be determined.

 Physical noise processes will be
detailed ahead but
 this indistinguishability of the signals
 without basis information is the clue for A and B to share
random bits over the Internet in a secure way. Physical noise has
been used before in fiber-optics based systems using $M$-ry levels
\cite{Mry} to protect information ($\alpha\eta$ systems). However,
the system proposed here is completely distinct from those
$\alpha\eta$ systems and it is related to the key distribution
system presented in \cite{barbosaKey}.


A brief description of protocol steps will be made, before a
theoretic-security analysis is shown and the system's limitations
discussed. It was said that if A and B start sharing a secret key
sequence ${\bf K_0}$ beforehand they may end up with a secure fresh
key sequence ${\bf K}$ much longer than ${\bf K_0}$ (${\bf K} \gg
{\bf K_0}$). Assume that ${\bf K_0}$ gives encoding information,
that is to say, which encoding ($V^{(0)}$ or $V^{(1)}$) is being
used at the $j^{\tiny \mbox{th}}$ emission. Assume that ${\bf
K}_0=k_1^{(0)},k_2^{(0)},...$ has a length $K_0$ and that the user A
has a physical random generator PhRG able to generate random bits
{\em and} noise in continuous levels. A generates a random sequence
${\bf K}_1=k_1^{(1)},k_2^{(1)},...k_{K_0}^{(1)}$ (say, binary
voltage levels)  and a sequence of $K_0$ noisy-signals $n$ (e.g.,
voltage levels in a continuum). The deterministic signal (carrying
recorded noise) ${\bf Y}_1=k_1^{(0)}\oplus f_1(
k_1^{(1)},n_1^{(1)}),k_2^{(0)}\oplus f_2(k_2^{(1)},n_2^{(1)}),...$
is then sent to B. Is B able to extract the fresh sequence ${\bf
K}_1$ from ${\bf Y}_1$? B applies ${\bf Y}_1 \oplus {\bf K_0}
=f_1(k_1^{(1)},n_1^{(1)}),f_2(k_2^{(1)},n_2^{(1)}),...f_N(k_N^{(1)},n_N^{(1)})$.
As B knows the encoding used and the signals representing bits 0 or
1 in a given encoding are easily identifiable:$\:\:\:$
$f_1(k_1^{(1)},n_1^{(1)})\rightarrow
k_1^{(1)},f_2(k_2^{(1)},n_2^{(1)})\rightarrow
k_2^{(1)},...f_N(k_N^{(1)},n_N^{(1)})\rightarrow k_N^{(1)}$. B then
obtains the new random sequence ${\bf K}_1$ generated by A.

Is the attacker also able to extract the same sequence ${\bf K}_1$?
Actually, this was a one-time pad with ${\bf K}_0$ with added noise
and, therefore, it is known that the attacker {\em cannot} obtain
${\bf K}_1$. The security problem arises for further exchanges of
random bits, e.g. if B wants to share further secret bits with A.


Assume that B also has a physical random generator PhRG able to generate random bits and noise in continuous levels. B wants
to send in a secure way a freshly generated key sequence ${\bf K}_2=k_1^{(2)},k_1^{(2)},...k_{K_0}^{(2)}$ from his PhRG to
A. B record the signals ${\bf Y}_2=k_1^{(1)}\oplus f_1( k_1^{(2)},n_2^{(2)}),k_2^{(1)}\oplus f_2(k_2^{(2)},n_2^{(2)}),...$
and sends it to A. As A knows ${\bf K}_1$  he(or she) applies ${\bf Y}_2 \oplus {\bf K_1}$ and extracts ${\bf K}_2$. A and B
now share the two new sequences ${\bf K}_1$ and ${\bf K}_2$. For speeding communication, even a simple rounding process to
the nearest integer would produce a simple binary output for the operation $f_j(k_j,n_j)$. The security of this process will
be shown ahead.



The simple description presented show a key distribution from A to B
and from B to A, with the net result that A and B share the fresh
sequences ${\bf K}_1$ and ${\bf K}_2$. These steps can be seen as a
first  distribution cycle.  A could again send another fresh
sequence ${\bf K}_3$ to B and so on. This repeated procedure
provides A and B with sequences ${\bf K}_1,{\bf K}_2,{\bf K}_3,{\bf
K}_4,...$. This is the basic key distribution protocol for the
system.

A last caveat should be made. Although the key sharing seems adequate to go without bounds, physical properties impose some
constraints and length limitations. Besides these limitations, the key sequences shared should pass key reconciliation and
privacy amplification steps \cite{Wolf} to establish security bounds to all possible E attacks. The length limitation arises
from the physical constraints discussed as follows.

A and B use PhRGs to generate physical signals creating the random
bits that define the key sequences ${\bf K}$ and the continuous
noise ${\bf n}$  necessary for the protocol. Being physical signals,
precise variables have to discussed and the noise source well
characterized. Interfaces will transform the physical signals onto
binary sequences adequate for Internet transmission protocols.
 Optical noise sources can be chosen for fast speeds. PhRGs have been
 discussed in the literature and even commercial ones are now
 starting to be available. Without going into details one could divide the PhRG in
 two parts, one generating random binary signals and another
 providing noise in a continuous physical variable (e.g., phase of a light
 field). These two signals are detected, adequately formatted and can be
 added.

 Taking the phase of a light field as the physical variable of interest,
 one could assume laser light in a coherent state with average number of photons $\langle  n \rangle$
 within one coherence time ($\langle  n \rangle=|\alpha|^2\gg 1$)
 and phase $\phi$.
Phases $\phi=0$ could define the bit 0 while $\phi=\pi$ could define
the bit 1.
 It can be shown \cite{barbosaKey} (see also ahead) that two non-orthogonal states with phases  $\phi_1$ and
 $\phi_2$ ($\Delta \phi_{12}=|\phi_1-\phi_2|\rightarrow 0$ and $\langle n \rangle \gg 1 $) overlap with (unnormalized) probability
\begin{eqnarray}
\label{pu} p_u \simeq e^{- (\Delta \phi_{12})^2 /2 \sigma_{
\phi}^2}\:\:,
\end{eqnarray}
where $\sigma_{\phi}=\sqrt{2/\langle n \rangle}$ is the standard
deviation measure for the phase fluctuations $\Delta \phi$. For
distinguishable states, $p_u \rightarrow 0$ (no overlap) and for
maximum indistinguishability $p_u=1$ (maximum overlap). With
adequate formatting $\phi_1 -\phi_2$ gives the spacing $\epsilon$
($\Delta \phi_{12} =\epsilon$) already introduced.  Eq. (\ref{pu})
with $\Delta \phi_{12}$ replaced by $\Delta \phi$   describes the
probability for generic phase fluctuations $\Delta \phi$ in a
coherent state of constant amplitude ($|\alpha|=\sqrt{\langle n
\rangle}=$constant) but with phase fluctuations.


The laser light intensity is adjusted by A (or B) such that
$\sigma_{\phi} \gg \Delta \phi$. This guarantees that the recorded
information in the files to be sent over the open channel is in a
condition such that the recorded light noise makes the two close
levels $\phi_1$ and $\phi_2$ indistinguishable to the attacker. In
order to avoid the legitimate user to confuse 0s and 1s in a {\em
single} basis, the light fluctuation should obey $\sigma_{\phi}\ll
\pi/2$. These conditions can be summarized as
\begin{eqnarray}
\label{condition}
 \frac{\pi}{2} \gg \sqrt{2/\langle n \rangle} \gg \Delta \phi
\:\:.
\end{eqnarray}
This shows that this key distribution system depends fundamentally
on physical aspects for security and not just on mathematical
complexity.

The separation between bits in the same encoding is easily carried
under condition  $ \pi/2 \gg \sqrt{2/\langle n \rangle}$. The
condition $\sqrt{2/\langle n \rangle} \gg \Delta \phi$ implies that
that set of bits 0--in encoding 0, and 1--in encoding 1 (set 1)
cannot be easily identifiable and the same happens with sets of bit
1--in encoding 0, and bit 0--in encoding 1 (set 2). Therefore, for
A, B and E, there are no difficulty to identify that a sent signal
is in set 1 or 2. However, E does not know the encoding provided to
A or B by their shared knowledge on the basis used. The question
``What is the attacker's probability of error in bit identification
without repeating a sent signal?'' has a general answer using
information theory applied to a binary identification of two states
 \cite{Helstrom}:
 The average probability of error in
identifying two states
 $| \psi_0\rangle$ and $| \psi_1\rangle$ is given by the Helstrom bound \cite{Helstrom}
\begin{eqnarray}
\label{Helstrom1}
 P_e=\frac{1}{2}\left[1-\sqrt{1-|\langle \psi_0|
\psi_1\rangle|^2}\: \right]\:\:.
\end{eqnarray}
Here $| \psi_0\rangle$ and $| \psi_1\rangle$ are coherent states of
light \cite{glauber} with same amplitude but distinct phases
\begin{eqnarray}
| \psi \rangle =|\alpha \rangle=| |\alpha|
e^{-i\phi}\rangle=e^{-\frac{1}{2}|\alpha|^2} \sum_n
\frac{\alpha^n}{\sqrt{n!}}| n \rangle \:, \end{eqnarray} defined at
the PhRG. $| \psi_0\rangle$ define states in encoding 0, where bits
0 and 1 are given by
\begin{eqnarray}
| \psi_0\rangle=\left\{
\begin{array}{c}
\hspace{3mm}| \alpha  \rangle, \:\:\:\mbox{for bit}\:\:\: 0,\:\:\:\mbox{and} \\
| -\alpha  \rangle, \:\:\:\mbox{for bit}\:\:\: 1\:\:,\end{array}
\right.
\end{eqnarray}
$| \psi_1\rangle$ define states in encoding 1, where bits 1 and 0
are given by
\begin{eqnarray}
| \psi_1\rangle=\left\{
\begin{array}{c}
| |\alpha |e^{-i \frac{\Delta \phi}{2}} \rangle, \:\:\:\mbox{for bit}\:\:\: 1,\:\:\:\mbox{and} \\
|  |\alpha| e^{-i \left( \frac{\Delta \phi}{2}+\pi \right)}
\rangle, \:\:\:\mbox{for bit}\:\:\: 0\:\:,\end{array} \right.
\end{eqnarray}
where $|\phi_0 - \phi_1|=\Delta \phi$. $|\langle \psi_0|
\psi_1\rangle|^2$ is calculated in a straightforward way and gives
\begin{eqnarray}
|\langle \psi_0| \psi_1\rangle|^2=e^{-2 \langle n \rangle \left[ 1-\cos \frac{\Delta \phi}{2}\right]}\:\:.
\end{eqnarray}
 For $\langle n \rangle \gg 1$ and
$\Delta \phi\ll 1$,
\begin{eqnarray}
|\langle \psi_0| \psi_1\rangle|^2\simeq e^{- \frac{\langle n \rangle }{4}\Delta \phi^2} \equiv e^{- \Delta \phi^2/\left( 2
\sigma_\phi^2\right)}\:\:,
\end{eqnarray}
where $\sigma_\phi=\sqrt{2/\langle n \rangle}$ is the irreducible
standard deviation for the phase fluctuation associated with the
laser field.

One should remind that in the proposed system the measuring
procedure is defined by the users A and B and {\em no} attack
launched by E can improve the deterministic signals that were
already available to him(her). Thus, the noise frustrating the
attacker's success, cannot be eliminated or diminished by
measurement techniques.

One should observe that each random bit defining the key sequence is
once sent as a message by A (or B) and then resent as a key
(encoding information) from B (or A) to A (or B). In both emissions,
noise is superposed to the signals. In general, coherent signal
repetitions implies that a better resolution may be achieved that is
proportional to the number of repetitions $r$. This improvement in
resolution is equivalent to a single measurement with a signal
$r\:\:\times$ more intense. To correct for this single repetition
$\langle n \rangle$ is replaced
 by $2\langle n \rangle$ in $|\langle \psi_0| \psi_1\rangle|^2$. The
final probability of error  results
\begin{eqnarray}
P_e=\frac{1}{2}\left[1-\sqrt{1-e^{-\frac{\langle n \rangle}{2}
\Delta \phi^2}} \:\right]\:\:.
\end{eqnarray}

This error probability  can be used to derive some of the proposed
system's limitations. The attacker's probability of success $P_s\:(
=1-P_e)$ to obtain the basis used in a single emission may be used
to compare with the a-priori starting entropy $H_{\small
\mbox{bases}}$ of the two bases that carry one bit of the message to
be sent (a random bit). If the attacker knows the basis, the bit
will also be known, with the same probability $\rightarrow 1$ as the
legitimate user.
\begin{eqnarray}
H_{\small \mbox{bases,bit}}=-p_0 \log p_0- p_1 \log p_1=1\:\:,
\end{eqnarray}
where $p_0$ and $p_1$ are the a-priori probabilities for each basis,
$p_0=p_1=1/2$, as defined by the PhRG. The entropy defined by
success events is $H_s=-P_s \log P_s$.  The entropy variation
$\Delta H=H_{\small \mbox{bases,bit}}-H_s$, statistically obtained
or leaked from bit measurements show the statistical information
acquired by the attacker with respect to the a-priori starting
entropy:
\begin{eqnarray}
\Delta H_{\small \mbox{bases}}=\left( H_{\small
\mbox{bases,bit}}-H_s \right)\:\:.
\end{eqnarray}
\begin{figure}
\centerline{\scalebox{0.6}{\includegraphics{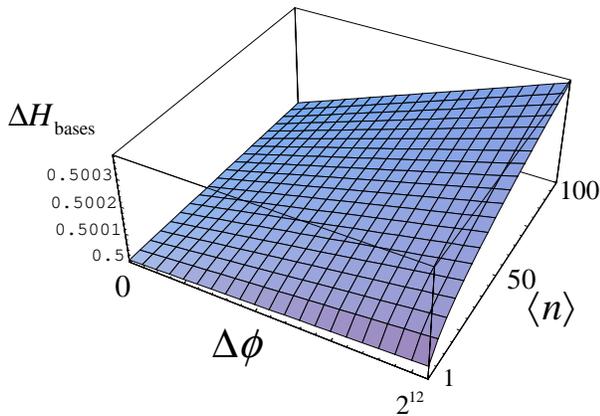}}} \caption{
$\Delta H_{\small \mbox{bases}}$ as a function of $\langle n
\rangle$ and $\Delta \phi$.} \label{delH}
\end{figure}
Fig. \ref{delH} shows $\Delta H_{\small \mbox{bases}}$ for some
values of $\langle n \rangle$ and $\Delta \phi$. Value $\Delta
H_{\small \mbox{bases}}=1/2$ is the limiting case where the two
bases cannot be distinguished. $\Delta H_{\small \mbox{bases}}$
deviations from this limiting value of $1/2$ indicates that some
amount of information on the basis used may potentially be leaking
to the attacker. It is clear that the attacker cannot obtain the
basis in a bit-by-bit process.
In order to be possible to obtain statistically a good amount of
information on a single {\em one} encoding used, $L$ should be given
by
\begin{eqnarray}
\label{length}
 L \times \left( \Delta H_{\small
\mbox{bases}}-\frac{1}{2}\right) \gg 1\:\:.
\end{eqnarray}
\begin{figure}
\centerline{\scalebox{0.45}{\includegraphics{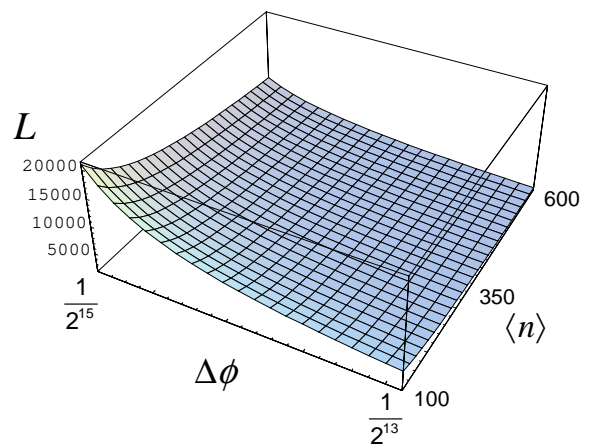}}}
\caption{Estimates for the minimum length of bits $L$ exchanged
between A and B that could give {\em one} bit of information about
the bases used to the attacker.} \label{leak}
\end{figure}
Fig. \ref{leak} shows estimates for $L$ for a range of values
$\langle n \rangle$ and $\Delta \phi$ satisfying $\label{L} L \times
\left( \Delta H_{\small \mbox{bases}}-\frac{1}{2}\right) =
1\:\:$($\Delta \phi$ is given in powers of 2, indicating bit
resolution for analog-to-digital converters).

It is assumed that error correction codes can correct for technical
errors in the transmission/reception steps for the legitimate users.
The leak estimate given by Eq. (\ref{length}) do not imply that the
information actually has leaked to the attacker.
 However, for security reasons, one takes for granted that this
deviation indicate a statistical fraction of bits acquired by the
attacker.

Privacy amplification procedures can be applied to the shared bits in order to reduce this hypothetical information gained
by the attacker to negligible levels \cite{Wolf}. These procedures are beyond the purposes of the present discussion but one
can easily accept that A and B may discard a similar fraction of bits to statistically reduce the amount of information
potentially leaked. Reducing this fraction of bits after a succession of bits are exchanged between A and B implies, e.g.,
that the number of bits to be exchanged will decrease at every emission. Eventually, a new shared key ${\bf K}_0$ has to
start the process again to make the system secure. Nevertheless, the starting key length $K_0$ was boosted in a secure way.
Without further procedures, the physical noise allowed  ${\bf K} \gg 10^3 {\bf K}_0$, a substantial improvement over the
classical one-time pad factor of 1. One may still argue that the ultimate security relies on ${\bf K}_0$'s length because if
${\bf K}_0$ is known no secret will exist for the attacker. This is also true but does not invalidate the practical aspect
of the system, because the ${\bf K}_0$ length can be made sufficiently long to frustrate any brute-force attack at any stage
of technology. Therefore, the combination of physical noise and complexity makes this noisy-one-time pad practical for
Internet uses.

Although the security of the process has been demonstrated,  one should also point to a fragility of the system (without a
privacy amplification stage) that has to be avoided when A and B are encrypting messages ${\bf X}$ between them. As it was
shown, knowledge of one sequence of random bits lead to the knowledge of the following sequence. This makes the system
vulnerable to know-plaintext attacks in the following way:
 E has a perfect record of both sequences
${\bf Y}_1$ and ${\bf Y}_2$ and tries to recover any bit sequence from them, ${\bf K}_2$, ${\bf K}_1$ or ${\bf K}_0$. E will
wait until A and B uses these sequences for encryption before trying to brake the system. A and B will encrypt a message
using a new shared sequence, ${\bf K}_1$ or ${\bf K}_2$. This message could be a plain-text, say ${\bf
X}=x_1,x_2,...x_{K_0}$ {\em known} to the attacker. Encrypting this message with say ${\bf K}_1$ in a noiseless way, gives
${\bf Y}=x_1 \oplus k_1^{(1)},x_2\oplus k_2^{(1)},...x_{K_0}\oplus k_{K_0}^{(1)}$. Performing the operation ${\bf Y}\oplus
{\bf X}$, E obtains ${\bf K}_1$. The chain dependence of ${\bf K}_j$ on ${\bf K}_{j-1}$ creates this fragility. Even
addition of noise to the encrypted file does not eliminate this fragility, because the attacker can use his/her knowledge of
${\bf X}$ --as the key-- to obtain ${\bf K}$--as a message. The situation is {\em symmetric} between B or the attacker: one
that knows the key (${\bf X}$ for E, and ${\bf K}$ for B) obtains the desired message (${\bf K}$ for E, and ${\bf X}$ for B)
 \cite{Ref2} .

In general, random generation processes are attractive to attackers and have to be carefully controlled. Well identifiable
physical components (e.g. PHRG) are usually a target for attackers that may try to substitute a true random sequence by
pseudo-random bits generated by a seed key under his/her control. Electronic components can also be inserted to perform this
task replacing the original generator; electric or electromagnetic signal may induce sequences for the attacker and so on.
In the same way, known-plaintext attacks also have to be carefully avoided by the legitimate users. The possibility of
further privacy amplification procedures to eliminate the known-plaintext attack presented is beyond the purposes of this
work.

Many protocols that use secret key sharing may profit from this
one-time pad booster system. For example, besides data encryption,
authentication procedures can be done by hashing of message files
with sequences of shared secret random bits. Challenge hand-shaking
may allow an user to prove its identity to a second user across an
insecure network.

As a conclusion, it has been shown that Internet users will succeed
in generating and sharing, in a fast way, a large number of secret
keys to be used in one-time-pad encryption as described. They have
to start from a shared secret sequence of random bits obtained from
a {\em physical} random generator hooked to their computers. The
physical noise in the signals openly transmitted is set to hide the
random bits. No intrusion detection method is necessary. Privacy
amplification protocols eliminate any fraction of information that
may have eventually obtained by the attacker. As the security is not
only based on  mathematical complexities but depend on physical
noise, technological advances will not harm this system. This is
then very different from systems that would rely entirely,  say, on
the difficulty of factoring large numbers in their primes. It was
then shown that by sharing secure secret key sequences, one-time pad
encryption over the Internet can be
practically implemented.  \\\\
$^*$E-mail:$\:\:$GeraldoABarbosa@hotmail.com


\end{document}